\documentclass{osa-article}

\journal{oe}

\begin{document}

\title{Proposal for a quantum traveling Brillouin resonator}

\author{Glen I. Harris\authormark{1,*}, Andreas Sawadsky\authormark{1}, Yasmine L. Sfendla\authormark{1}, Walter W. Wasserman\authormark{1}, Warwick P. Bowen\authormark{1}, Christopher G. Baker\authormark{1}}

\email{\authormark{*}g.harris2@uq.edu.au}
\address{\authormark{1}ARC Centre of Excellence for Engineered Quantum Systems, School of Mathematics and Physics, University of Queensland, St Lucia, QLD 4072, Australia.}%

\begin{abstract}
Brillouin systems operating in the quantum regime have recently been identified as a valuable tool for quantum information technologies and fundamental science. However, reaching the quantum regime is extraordinarily challenging, owing to the stringent requirements of combining low thermal occupation with low optical and mechanical dissipation, and large coherent phonon-photon interactions.
Here, we propose an on-chip liquid based Brillouin system that is predicted to exhibit ultra-high coherent phonon-photon coupling with exceptionally low acoustic dissipation. The system is comprised of a silicon-based ``slot'' waveguide filled with superfluid helium. This type of waveguide supports optical and acoustical traveling waves, strongly confining both fields into a subwavelength-scale mode volume. It serves as the foundation of an on-chip traveling wave Brillouin resonator with a single photon optomechanical coupling rate exceeding $240$~kHz. Such devices may enable applications ranging from ultra-sensitive superfluid-based gyroscopes, to non-reciprocal optical circuits. Furthermore, this platform opens up new possibilities to explore quantum fluid dynamics in a strongly interacting condensate.
\end{abstract}

\section{Introduction}

\begin{figure}
\includegraphics[width=.9\columnwidth]{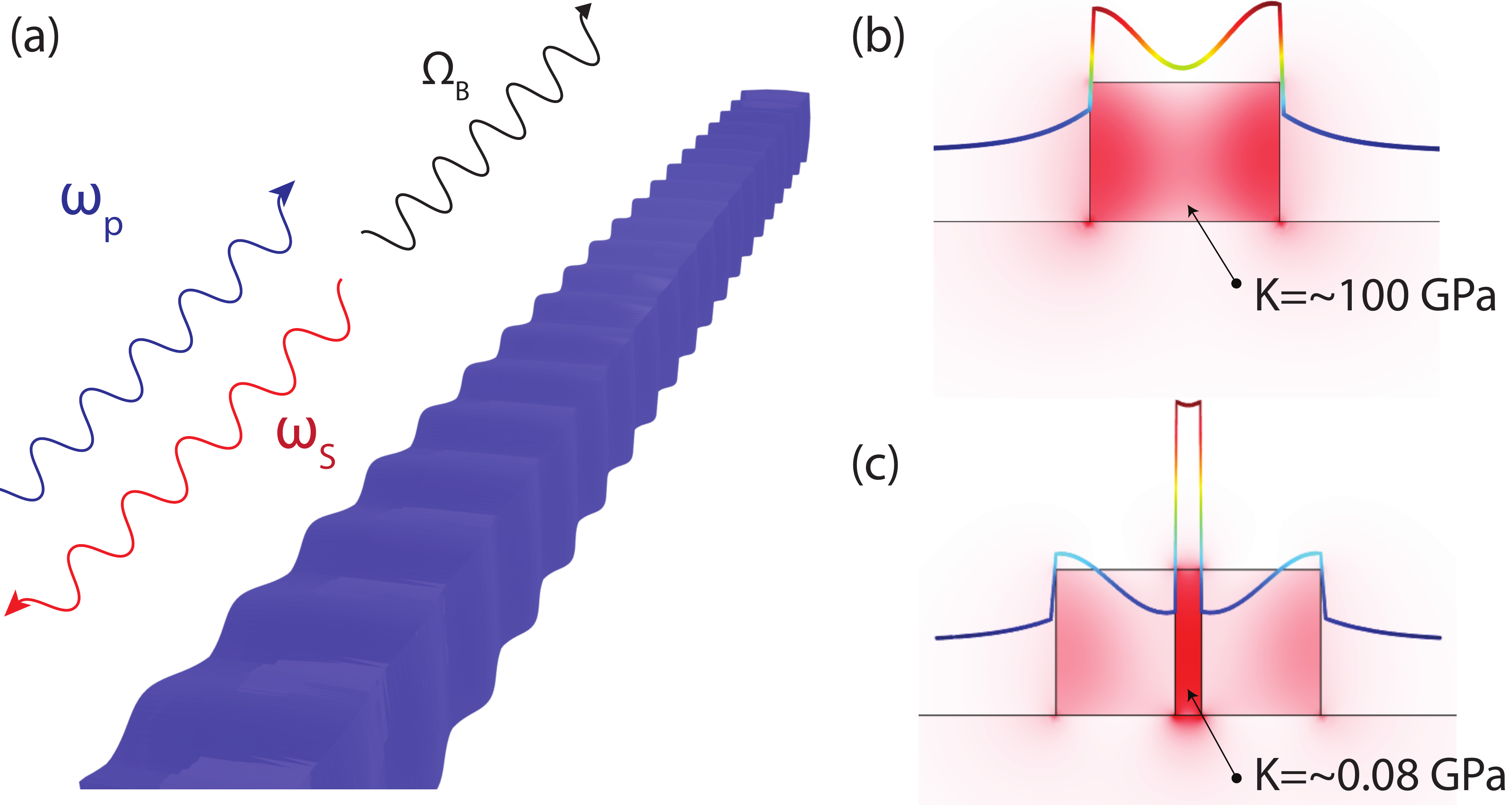}
\caption{\label{Figure1} (a) Schematic description of the backward Brillouin scattering process: a pump photon $\omega_p$ scatters off a moving refractive index grating creating a red-shifted Stokes photon $\omega_S$ and forward propagating acoustic phonon $\Omega_B$. (b) Optical field distribution inside a dielectric ridge waveguide. $K$ refers to the bulk modulus, on the order of hundreds of gigapascals for many solids. (c) Field distribution in a slot waveguide, where the peak field intensity is located within the mechanically compliant ($K=0.08$ GPa) superfluid filling the void.}
\end{figure}

For many decades, the Brillouin interaction was viewed as a weak nonlinear process, only appearing at high optical powers which were large enough to reveal the bulk electrostrictive properties of a material. However, this perspective was challenged in the early 2000's after it was observed that co-localising light and sound into the confined geometry of a photonic crystal fiber resulted in strong Brillouin scattering at relatively low optical powers\cite{Dainese_NatPhys_06}. It was further recognised that confining optical fields to sub-wavelength waveguides could also enable radiation pressure forces through boundary deformations, such as those illustrated in Fig. \ref{Figure1}(a), resulting in an additional optical force that could be tailored through waveguide design\cite{rakich_giant_2012}. 
These insights, combined with accessible nano-fabrication technologies, led to an unprecedented growth in nanoscale photonics research across a wide variety of platforms\cite{eggleton_brillouin_2019}, including silica photonic crystal fibers\cite{Dainese_NatPhys_06}, suspended silicon beams\cite{rakich_giant_2012}, hybrid silicon-chalcogenide waveguides\cite{Zhou_OptExp_19}, encapsulated silicon nitride waveguides\cite{Gyger_PRL_20}, and calcium fluoride resonators \cite{grudinin_brillouin_2009}. These platforms have been used to demonstrate a number of promising Brillouin based applications, such as optical delay and memory\cite{Merklein_NatComm_17, Zhu_Science_07}, non-reciprocal optical modulation \cite{Kittlaus_NatPhot18}, microwave-photonic filtering \cite{marpaung_low-power_2015}, ultra-precise rotation sensing \cite{Lai_NatPhot_20}, optical amplification \cite{Kittlaus_NatPhot16} and narrow linewidth Brillouin-based lasers \cite{gundavarapu2019sub, otterstrom_silicon_2018}. However, the vast majority of these demonstrations only leverage the classical aspects of the Brillouin interaction, and are typically limited by either excess thermal noise, weak interaction strengths or large optical and mechanical dissipation rates. It is expected that by surpassing these limitations and harnessing the quantum mechanical aspects of the interaction, Brillouin systems will play a key role in emerging quantum technologies and fundamental science \cite{SafaviNaeini_Optica_19, Rakich_NJP_18, kashkanova2017superfluid, Enzian_Optica_19}. 
For example, a Brillouin system may act as delay line\cite{Zhu_Science_07} or optical router\cite{Kittlaus_NatPhot18,Sohn_NatPhot_2018} for optical quantum communication, or even provide a quantum interface between superconducting circuits ($\sim$GHz) and optical quantum communication channels ($\sim$THz) \cite{SafaviNaeini_Optica_19}.
Indeed, such components are considered prerequisites for a ``quantum internet'' to distribute quantum resources for computation, communication and metrology across the globe\cite{Kimble_Nat_08}.

In systems that rely on acoustic and optical resonant enhancement, a simple figure of merit to benchmark quantum capabilities is the single photon cooperativity $C_0 = \frac{4 g_0^2}{\kappa \Gamma}$ \cite{bowen2015quantum}. It relates the strength of the light-sound coupling rate $g_0$ to the optical and acoustic decay rates of the system, labelled here as $\kappa$ and $\Gamma$, respectively. For context, with $C_0 > 1$ in a system with low thermal occupancy, a single intracavity photon will interact with a single phonon to hybridise into a new quasi-particle. Furthermore, this condition ($C_0 >1$) also signifies that the Brillouin lasing threshold can be reached with just a single intracavity photon. 
Surprisingly, despite substantial gains over the last decade to increase light-sound coupling rates and attain higher cooperativities, state-of-the-art Brillouin based systems typically have $C_0 < 0.1$ \cite{wiederhecker_brillouin_2019, VanLaer_PRA_2016}.  It should be noted that in non-resonant systems, such as continuous waveguides, the single photon cooperativity is not a common figure-of-merit. Instead, the Brillouin gain $G_B$ is often used, which quantifies the amplification of the Stokes beam for a given pump power and waveguide length. Indeed, these two figures-of-merit are intrinsically linked, which can be intuited by considering the lasing threshold of a waveguide that is deformed into a closed path (i.e. an optical and acoustic cavity). Expressions that convert between these figures-of-merit can be found in Ref.\cite{VanLaer_PRA_2016} and Ref.\cite{wiederhecker_brillouin_2019}.

Recently, superfluid helium has successfully demonstrated its utility as a quantum optomechanical system \cite{harris_laser_2016, mcauslan_microphotonic_2016, Shkarin_PRL19}, owing to its ultra-low electromagnetic and acoustic dissipation \cite{lorenzo_superfluid_2014, souris2017ultralow, kashkanova2017superfluid}. Superfluid helium conforms to any electromagnetic cavity, which helps provide good overlap between the acoustic and optical fields, a requirement for high optomechanical coupling rates. Furthermore, its negligible optical absorption, vanishing viscosity and large mechanical compliance make it a particularly desirable substance for Brillouin scattering. However, one limitation  of superfluid helium  lies in its minute refractive index ($n_\mathrm{He} \simeq 1.029$), which makes it difficult to use it to confine light and thus co-localize light and sound in a small interaction volume. This can be overcome to some degree by confining the superfluid to a micron-scale Fabry-Perot cavity \cite{kashkanova2017superfluid, Shkarin_PRL19}, or by using a higher index waveguiding structure (such as a microdisk or microtoroid) to confine the light, with coupling to the superfluid afforded by the evanescent component of the light which extends outside the resonator \cite{baker_theoretical_2016, harris_laser_2016, sachkou_coherent_2019, he_strong_2019}. This approach however generally precludes simultaneous optical and mechanical confinement to wavelength-scale interaction volumes, limiting the light-sound coupling rates.

To overcome this, we propose an on-chip liquid based Brillouin system that is predicted to exhibit ultra-high light-sound coupling with exceptionally low acoustic dissipation and low thermal occupancy. It relies on a slot waveguide resonator geometry\cite{almeida_guiding_2004, xu_experimental_2004}, where the optical field maximum is localized within a narrow channel which is filled with superfluid helium, as illustrated in Fig. \ref{Figure1}(c) and discussed in detail below. Thanks to the combination of small mode volume and superfluid helium's high mechanical compliance, we calculate that such devices may enable optomechanical coupling rates on the order of $g_0/2\pi\sim 250$ kHz and single photon cooperativities exceeding unity in a backward Brillouin scattering regime. A further benefit of this approach is afforded by the low bulk modulus of liquid helium relative to most Brillouin active solids (see Fig.\ref{Figure1}(b) and Fig.\ref{Figure1}(c) for comparison). This results in a small Brillouin shift of only $\sim400$~MHz in liquid helium, and enables the simultaneous resonant enhancement of the pump and Stokes beams, independent of device size. Indeed, this newly identified process, called counter-modal Brillouin scattering, has recently been shown to greatly enhance the strength of the Brillouin interaction in micron-scale cavities coated with superfluid helium \cite{he_strong_2019}. This is in contrast to intra-modal or inter-modal Brillouin scattering process where the device's size and mode structure must be carefully engineered to ensure the energy and momentum matching conditions are satisfied \cite{wiederhecker_brillouin_2019}.

\begin{figure*}
\centering\includegraphics[width=0.9\textwidth]{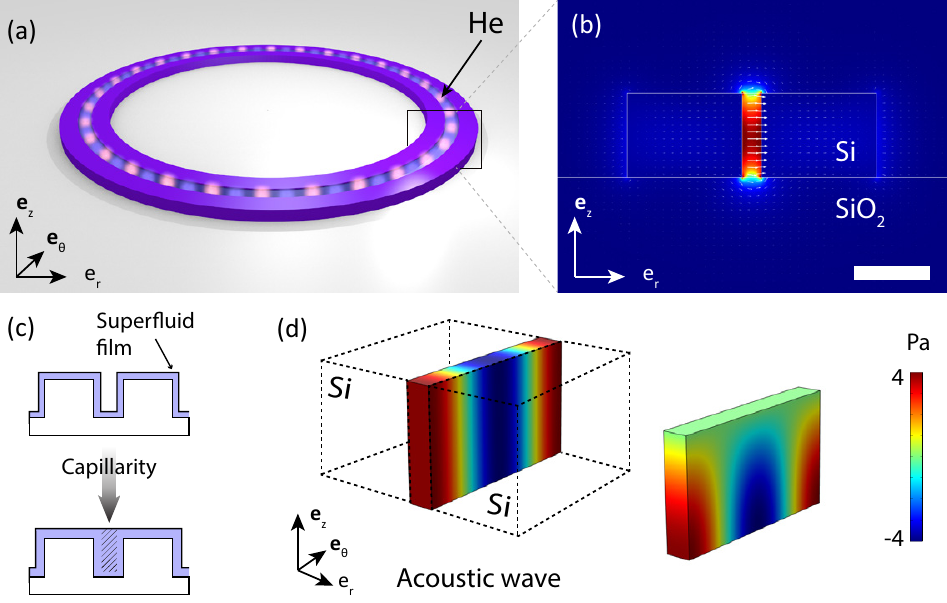}
\caption{\label{Figure2}(a) Illustration of the proposed device. The silicon is shown in purple and the underlying silica in white. Superfluid helium fills the narrow slot between the silicon waveguides, where the optical field intensity is largest. (b) Cross-section of the electric field intensity distribution $\left|E\right|^2$ for the transverse electric (TE) whispering gallery mode (WGM) guided within the slot waveguide resonator. White arrows represent the electric field orientation predominantly along $\textbf{e}_r$. Scale bar corresponds to 200 nm. (c) Illustration of the slot filling with superfluid by capillarity for sufficiently thick (exceeding $\sim2$ nm) films. (d) Finite element simulation of the zero-point pressure excursion due to the Brillouin density wave inside the superfluid, in the case of fixed (left) and free (right) boundary conditions for superfluid flow at the top of the slot. Zero-point pressure excursions are typically on the order of a few Pascal, with values provided here for a 50 nm wide slot. Only one acoustic wavelength is shown here; this pressure pattern is repeated along the entire slot circumference. }
\end{figure*}

\section{Proposed implementation}

\begin{figure*}
\centering\includegraphics[width=0.9\textwidth]{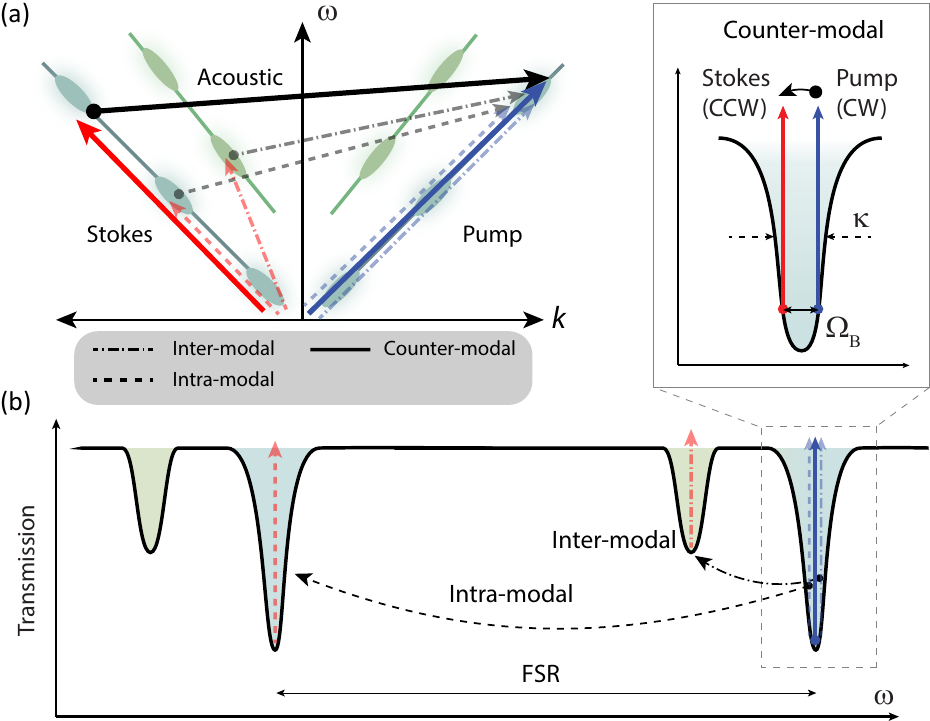}
\caption{\label{Figure4} Backward Brillouin scattering schemes in microresonators. (a) Optical dispersion diagram. Light blue and green lines represent the dispersion branches for different optical modes, and ellipses the discrete optical resonances of differing azimuthal order. Solid, dashed and dashed-dotted arrows respectively represent the counter-modal, intra-modal and inter-modal backward Brillouin scattering cases. (b) Frequency space illustration of the intra-modal, inter-modal and counter-modal (inset) scattering cases, where $\kappa$ is the optical linewidth. WGMs of a same family, separated by a free spectral range (FSR)---\textit{i.e.} differing by one azimuthal node---are shown with the same color shading. Inset: illustration of counter-modal Brillouin scattering~\cite{he_strong_2019}, where pump and Stokes beams are hosted by degenerate clockwise and counter-clockwise whispering gallery modes of identical azimuthal order.  }
\end{figure*}

Figure \ref{Figure2}(a) shows an illustration of the proposed implementation. Light is confined inside a slot waveguide ring resonator, built from a silicon on insulator (SOI) wafer.
These slot resonators rely on the large normal electric field discontinuity at the interface of a high refractive index material to achieve waveguiding with a substantial fraction of the energy located within the low-index empty slot \cite{almeida_guiding_2004, xu_experimental_2004}, as plotted in Fig. \ref{Figure2}(b).
Because of the small mode volume and strong light-matter interaction they can provide, resonators based on this geometry have been successfully leveraged for a broad range of applications  including on-chip gas \cite{robinson_-chip_2008} and bio-sensing \cite{fcarlborg_packaged_2010}, as well as optomechanics \cite{li_ultrahigh-frequency_2010} and electro-optic modulation~\cite{gould_silicon-polymer_2011}.
When placed in a cryostat in the presence of helium below the superfluid transition temperature, the device will be coated with a thin, self-assembling superfluid film~\cite{harris_laser_2016,mcauslan_microphotonic_2016,sachkou_coherent_2019, he_strong_2019}, which can fill the slot volume with superfluid helium through capillary action \cite{yang_coupling_2016,kashkanova_optomechanics_2017}, as illustrated in Fig. \ref{Figure2}(c). 
This will occur when the added van der Waals potential energy in the hatched area in Fig. \ref{Figure2}(c) is more than offset by the decrease in surface tension energy due to the reduced surface area.  Calculations (not shown here) show that for this proposed implementation, the filling transition occurs for film thicknesses in excess of approximately 2 nm. Hence the slot resonator topology provides a simple, robust, and alignment-free way to co-localise light and sound to wavelength scale mode volumes.

The Brillouin interaction occurs due to the electrostrictive coupling between the optical field and travelling acoustic waves in the superfluid causing periodic density and refractive index fluctuations~\cite{boyd2003nonlinear,kashkanova2017superfluid,he_strong_2019,baker_photoelastic_2014}. (Radiation pressure forces exerted on the silicon slabs are not considered here \cite{li_ultrahigh-frequency_2010, sarabalis_guided_2016}).
Density waves in a superfluid confined to a narrow channel are named `fourth-sound', which describes a wave in which the superfluid component is free to oscillate while the normal fluid component is viscously clamped~\cite{atkins_third_1959, tilley_superfluidity_1990}. However in the limit of millikelvin temperatures at which these experiments take place\cite{he_strong_2019, kashkanova2017superfluid}, these reduce to simple density waves, called `first-sound' in the superfluid context~\cite{tilley_superfluidity_1990}.
Thanks to the superfluid's inviscid and irrotational nature, these can be modelled as sound waves of an ideal gas~\cite{forstner_modelling_2019}.
Confinement of the acoustic wave within the slot is provided by the solid silica and silicon boundaries at the bottom and lateral sides. The confinement at the top of the slot  can be provided by two different means. The slot could be sealed shut, for example via wafer bonding, resulting in a fixed boundary condition for fluid displacement (which corresponds to a free pressure boundary condition~\cite{forstner_modelling_2019}). Alternatively, the slot can remain open at the top, in which case the confinement in the vertical direction is provided by the surface tension at the superfluid interface. Owing to the large difference in stiffness between surface tension and bulk modulus,
this results in an effective free boundary condition for fluid displacement, \textit{i.e.} a fixed pressure boundary condition~\cite{forstner_modelling_2019}, as illustrated in Fig. \ref{Figure2}(d).  While the fluid in the slot is in this case connected to the rest of the chip via a thin superfluid film, acoustic losses are suppressed by the large acoustic impedance mismatch~\cite{poulton_acoustic_2013} between the speed of sound in the bulk helium within the slot ($c_1=238$ m/s) and the low speed of surface waves in the thin film elsewhere on the chip ($c_3\simeq 1$ m/s)\cite{sachkou_coherent_2019,baker_theoretical_2016}. As we will see in the following, both approaches yield  very similar performance. The ring cavity thus forms  a high finesse cavity for both optical and superfluid acoustic waves.

The proposed dimensions of the device and physical parameters used for the performance estimations in the following are provided in Table \ref{tableexperimentalparameters}. We model a 20 $\mu$m diameter slot ring resonator, with a slot width ranging from 5 nm to 150 nm, which corresponds to an extremely minute optical mode volume on the order of the optical wavelength $\lambda_{\mathrm{opt}}$ cubed. As a general rule, miniaturization is desirable, as it enables a large increase in the Brillouin interaction strength due the efficient co-localization of light and sound. So far this goal has been hampered by the fact that the usual energy and momentum matching criteria  generally cannot be met in miniaturized devices. Indeed, the frequencies of the optical pump and the acoustic wave must be such that the pump \emph{and} the frequency-shifted Stokes beam match a resonant optical mode of the device~\cite{wiederhecker_brillouin_2019}. For backward Brillouin scattering in solid resonators, this requirement is typically achieved through  two distinct approaches. The first, \emph{intra-modal} scattering, relies on pump and scattered fields being resonant with adjacent WGMs of the same mode family, removed by one or multiple free spectral ranges\cite{wiederhecker_brillouin_2019, li_microresonator_2017}, as illustrated by the dashed line in Figs. \ref{Figure4}(a) and \ref{Figure4}(b).
Alternatively this can be achieved by having pump and scattered fields resonant with different spatial modes, a process named \emph{inter-modal} Brillouin scattering \cite{wiederhecker_brillouin_2019,eggleton_brillouin_2019}, as illustrated by the dash-dot line in Figs. \ref{Figure4}(a) and \ref{Figure4}(b).

Intra-modal scattering imposes limitations on the device's ultimate miniaturization, as the FSR cannot exceed the Brillouin shift. Moreover, both intra-modal and inter-modal scattering schemes only function with devices of very specific dimensions, such that the spacing between the chosen optical modes exactly matches the Brillouin shift, as illustrated in Fig. \ref{Figure4}(b), a condition which becomes particularly challenging in miniaturized devices with sparse optical spectra. Here we propose to leverage superfluid helium's large mechanical compliance (the bulk modulus of helium $K_{\mathrm{He}}=8$ MPa, \emph{vs} tens of GPa for most solids) to overcome this limitation. This large compliance leads to a low speed of sound $c$, and correspondingly  to a small Brillouin shift $\Omega_B/2\pi=2 \,c\, n_{\mathrm{eff}}/ \lambda_{\mathrm{opt}} \sim 400$ MHz, close to 30 times smaller than that in bulk silica. This small shift means that for WGMs with optical quality factors below $\sim 5\times 10^5$ (whose optical linewidth > 400 MHz) both the pump and Stokes beam can naturally fall within a single optical resonance, as illustrated in Fig. \ref{Figure4}(b). This enables \emph{counter-modal} Brillouin scattering between degenerate, counter-propagating (clockwise and counter-clockwise) whispering gallery modes of same azimuthal order $m_{\mathrm{opt}}$, as demonstrated in our recent work with superfluid third-sound waves \cite{he_strong_2019}. This triple resonant enhancement (pump, Stokes and mechanical wave) is thus automatically satisfied independently of device size, while enabling an electrostrictive coupling ($g_0/2\pi$) in excess of $10^5$ Hz to be achieved, as detailed below.

The present approach presents two key distinguishing features compared to the work outlined in Refs. \cite{kashkanova2017superfluid, shkarin_quantum_2019}, which employed the electrostrictive coupling between light and bulk superfluid helium contained within a fiber cavity.
First, the approximatively $10^4$ times smaller mode volume leads to a $\sim10^4$ increase in the per photon cooperativity\cite{bowen2015quantum} $C_0=4g_0^2/(\kappa \Gamma)$ and an associated four orders of magnitude decrease in the lasing threshold, all else being equal.
Second, in refs \cite{kashkanova2017superfluid, shkarin_quantum_2019} both photons and phonons are confined within a Fabry-Perot cavity, leading to standing optical and acoustic fields. In contrast, our implementation couples travelling optical and acoustic fields, a key requirement for non-reciprocity \cite{Kittlaus_NatPhot18}.

\begin{table}
\centering
    \begin{tabular}{l|c|c|c}
        \hline
        Parameter                & Symbol       & Value  & Units       \\
           \hline
           \hline
               Ring outer radius               & $R$                & $ 10 $  & $\mu$m  \\  
               Slot height & $h$ & 220 & nm \\
    WGM azimuthal number        & $m_{\mathrm{opt}} $                & $186 $ & -  \\ 
    Mechanical azimuthal number        & $m $                & $372 $ & - \\ 
    Speed of first sound 
    & $c_1$ & $238$  & m/s  \cite{donnelly_observed_1998} \\ 
        Refractive index of superfluid $^4$He  & $n_{\mathrm{He}}$ & $1.029$  & -  \cite{donnelly_observed_1998} \\ 
          Superfluid $^4$He density  & $\rho_{\mathrm{He}}$ & $145$  & kg/m$^3$  \cite{donnelly_observed_1998} \\ 
           Superfluid $^4$He bulk modulus  & $K_{\mathrm{He}}$ & $8.21$  & MPa  \\ 
    Brillouin mode zero-point motion                   & $x_{\mathrm{ZPF}} $   & $9.5 \times 10^{-15}$  & m  \\
      Brillouin shift (typical) & $\Omega_B/2\pi$& $\sim 400$& MHz\\
        \hline
    \end{tabular}
\caption{Values of the parameters used in the numerical simulation. }
\label{tableexperimentalparameters}
\end{table}

\section{Calculation of the coupling strength}
We compute the strength of the single photon electrostrictive coupling rate $g_{0,\mathrm{es}}$ between the optical and acoustic fields\cite{wiederhecker_brillouin_2019}, as a function of the slot width.
It is given by \cite{he_strong_2019, kashkanova2017superfluid}:
\begin{equation}
g_{0,\mathrm{es}}=\frac{\omega}{2}\frac{\int \gamma_e \, \tilde{\epsilon}_{v}\left(\vec{r}\right) E_p\left(\vec{r}\right) \, E_s\left(\vec{r}\right) \mathrm{d}^3 \vec{r} }{\sqrt{\int \varepsilon_r E_p\left(\vec{r}\right) \,\mathrm{d}^3 \vec{r}}\sqrt{\int \varepsilon_r E_s\left(\vec{r}\right) \,\mathrm{d}^3 \vec{r}}},
\label{Eq_g_0es2}
\end{equation}
where $E_p$ and $E_s$ respectively refer to the pump and Stokes electric fields, $\varepsilon_r$ is the relative dielectric permittivity, $\tilde{\epsilon}_v = \frac{\delta V_{zp}}{V} = -\delta P_{zp}/K_{\mathrm{He}} = -\frac{\delta\rho_{zp}}{\rho}$ is the zero-point volumetric strain caused by the Brillouin wave in the superfluid, and 
\begin{equation}
\gamma_e=\left(\rho\, \frac{\partial\varepsilon_r}{\partial \rho}\right)_{\rho=\rho_0}=\left(\varepsilon_r-1\right)\left(\varepsilon_r+2\right)/3
\label{Eqelectrostrictivecoefficient}
\end{equation}
is the electrostrictive constant of the material \cite{boyd2003nonlinear}, derived from the Clausius-Mossotti relation for an isotropic dielectric. Thus $\gamma_e \, \tilde{\epsilon}_{v}$ corresponds to the zero-point permittivity fluctuations of the material due to the Brillouin wave, which is normalized such that its integrated strain energy density is equal to a half quantum of energy, \textit{i.e.}  $\int \frac{1}{2} K_{\mathrm{He}} \tilde{\epsilon}_{v}^2\left(\vec{r}\right)  \,\mathrm{d}^3 \vec{r}=\hbar \Omega_B /2$. Here $K_{\mathrm{He}} = \rho_{\mathrm{He}}\,c_{1}^2 = 8.21\times 10^6$ Pa is the bulk modulus of superfluid helium. For backward Brillouin scattering with identical azimuthal order pump and Stokes waves, as is the case here, and superfluid helium for which $\varepsilon_{\mathrm{sf}} = 1.058\approx 1$, Eq.~(\ref{Eq_g_0es2}) simplifies to:
\begin{equation}
g_{0,\mathrm{es}}=\frac{\omega}{2} \left(\varepsilon_{\mathrm{sf}} -1 \right)\frac{\int_{\mathrm{slot}}  \tilde{\epsilon}_{v}\left(\vec{r}\right) E^2\left(\vec{r}\right) \, \mathrm{d}^3 \vec{r} }{\int_{\mathrm{all}} \varepsilon_r\left(\vec{r}\right) E^2\left(\vec{r}\right) \,\mathrm{d}^3 \vec{r}},
\label{Eq_g_0es1}
\end{equation}
where the top integral describing the overlap between the optical field and superfluid strain field is carried out over the slot interaction region, while the bottom integral normalizing the optical field is carried out over all space.

\begin{figure*}
\centering\includegraphics[width=\textwidth]{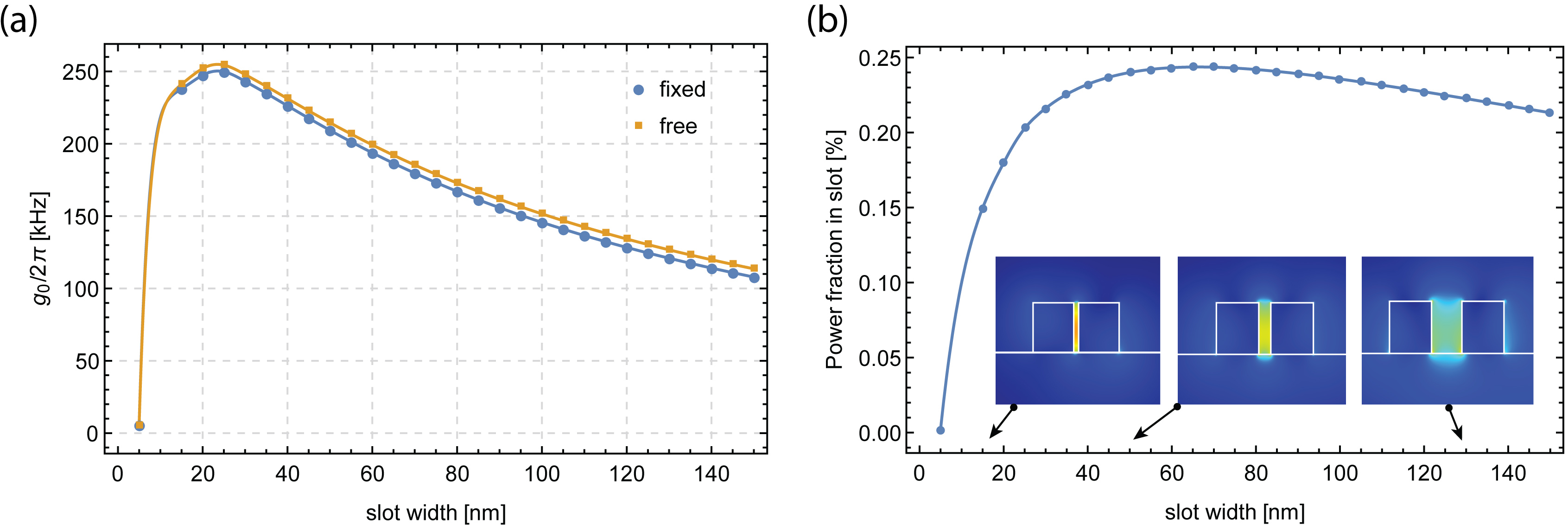}
\caption{\label{Figure3}(a) Single photon optomechanical coupling rate $g_0$ as a function of slot width, for fixed (blue) and free (orange) boundary conditions for the acoustic wave (see Fig. \ref{Figure2}(c)). (b) Fraction of the WGM electromagnetic energy contained  within  the slot, as a function of slot width. Insets show the electric field intensity distribution $\left|E\right|^2$ in the resonator for slots widths of 15 nm; 50 nm and  130 nm. Resonator diameter and slot height are kept fixed at 20 $\mu$m and 220 nm respectively, see Table \ref{tableexperimentalparameters}.  }
\end{figure*}

We numerically calculate $g_{0,\mathrm{es}}$, through Eq. (\ref{Eq_g_0es1}), as a function of the slot width.  Optical field and superfluid zero-point pressure fluctuations are computed using the finite element tool Comsol Multiphysics\cite{baker_theoretical_2016, forstner_modelling_2019}, as shown in Figs. \ref{Figure2}(b) and \ref{Figure2}(d). We choose an optical resonance near $\lambda_{\mathrm{opt}}=1550$ nm and select the corresponding superfluid acoustic eigenmode with azimuthal order $m=2 m_{\mathrm{opt}}$. This ensures the energy and momentum matching conditions for Brillouin scattering are satisfied, which in WGM resonators are given by $\omega_{\mathrm{pump}}=\omega_{\mathrm{Stokes}}+\Omega_B$ and $m=m_{\mathrm{pump}} \pm m_{\mathrm{Stokes}}$, where + and - respectively correspond to the backward and forward scattering processes.

The result is shown in Fig. \ref{Figure3}(a). The coupling rate $g_{0, \mathrm{es}}$ initially increases with decreasing slot width, up to a value of  $\sim 2\pi\times  250$ KHz (more than one order of magnitude larger than the radiation pressure coupling achieved with superfluid coated microresonators \cite{he_strong_2019}), before abruptly decreasing as the slot width is further reduced below  $\sim 20$ nm. This qualitative behaviour arises due to two competing trends. On one hand, the zero-point  strain and pressure fluctuations monotonically increase with decreasing slot width, as the zero-point energy is distributed over a smaller volume. On the other hand, this trend is offset by the diminishing fraction of the optical power located within the slot which effectively participates in the Brillouin interaction. This fraction peaks at $\sim 25$ \% for a slot width of 60 nm, before decreasing to zero for a vanishing slot width, as illustrated in Fig. \ref{Figure3}(b).

The different boundary conditions have minimal impact on the expected $g_{0,es}$, with peak values in both cases on the order of $2\pi\times 250$ kHz, and $g_{0,es}/2\pi$ in excess of 150 kHz for slot widths ranging from approximately 5 nm to 150 nm. Indeed, while the fixed pressure boundary condition has a weaker overlap with the optical field, it exhibits higher zero-point pressure excursions due to its reduced mode volume (see Fig. \ref{Figure2}(d)), leading overall to a similar coupling strength.
It is important to note that the Brillouin shift $\Omega_B/2\pi=2\, c\, n_{\mathrm{eff}}/ \lambda_{\mathrm{opt}}$ depends on the effective index $n_{\mathrm{eff}}$ of the guided optical mode---which itself is dependent upon slot and silicon ridge widths and can take values between the free space value of 1 and the bulk silicon value of 3.48 dependent on device geometry. Because of this, the Brillouin frequency is here not solely determined by the bulk properties of superfluid helium~\cite{st._peters_brillouin_1970, kashkanova2017superfluid}, and tunable Brillouin shifts ranging from $\sim300$ MHz to 1 GHz can be accessed for Telecom wavelength light, providing a degree of tunability that is not as accessible in other systems.

\section{Discussion}
The on-chip photonic implementation proposed here is compatible with robust optical packaging for cryogenic operation~\cite{mckenna_cryogenic_2019}, which greatly simplifies experimental design by alleviating the need for cryogenic alignment of optical elements. Furthermore, thanks to the higher acoustic frequency of first sound compared to third sound~\cite{he_strong_2019}, the Brillouin wave would be in or near its quantum ground-state when thermalized in a dilution refrigerator~\cite{shkarin_quantum_2019}. Efficient thermal anchoring is provided by the large contact area with the surrounding silicon, and through superfluid helium's large thermal conductivity. This stands in contrast to competing systems with high single photon cooperativity, such as one-dimensional photonic crystal cavities, which suffer from greatly increased dissipation rates for even near-unity intracavity photon numbers, due to their reduced thermal anchoring \cite{ren_twodimensional_2019}.

Calculation of the single photon cooperativity $C_0=\frac{4 g_0^2}{\kappa \Gamma}$ requires an estimate for the acoustic damping rate $\Gamma$~\cite{bowen2015quantum}. Mechanical quality factors as high as $10^8$, limited by $^3$He impurities, have been demonstrated in bulk superfluid $^4$He resonators~\cite{de_lorenzo_ultra-high_2017}, and on the order of $10^5$ in helium filled fiber cavities~\cite{kashkanova2017superfluid, shkarin_quantum_2019}. In the latter case, the acoustic dissipation was ultimately limited by acoustic energy transfer into the fibers from the superfluid phonons impinging at normal incidence onto the fiber cavity end-facets. In contrast, in the slot waveguide ring resonator, the phonons would constantly be reflected at grazing incidence, forming a superfluid acoustic whispering gallery mode and potentially alleviating that loss mechanism. For acoustic quality factors ranging from $10^5$ to $10^8$, large single photon optomechanical cooperativities exceeding unity could be achieved in a travelling Brillouin system: with $C_0$ ranging from $6 \times 10^{-2}$ to 60, assuming an optical linewidth of $\kappa/2\pi=1$ GHz. For even a modest acoustic $Q=10^4$, the lasing threshold for which the Brillouin gain exceeds the acoustic dissipation occurs at approximately one hundred nanowatts.

If the optical linewidth can be reduced to values below $400$~MHz, then the system proposed here will be operating in the `resolved sideband' regime, where  $\Omega > \kappa$ \cite{bowen2015quantum}. This allows one to use the modeshape of the optical cavity to selectively enhance or suppress Stokes or anti-Stokes transitions, in both clockwise and counter-clockwise directions. By considering the Hamiltonian picture of the Brillouin interaction\cite{he_strong_2019, VanLaer_PRA_2016}, it can be shown that this confers the ability to perform a quantum mechanical "state-swap" between the optical and mechanical modes (when enhancing the anti-Stokes process)\cite{Enzian_Optica_19}, or enables the generation of two-mode squeezing/entanglement between optical and mechanical degrees of freedom (when enhancing the Stokes process). Indeed, these techniques have been widely investigated in low-frequency mechanical systems driven by radiation pressure\cite{bowen2015quantum} (i.e. cavity optomechanics), with recent high-profile demonstrations of coherent state-swaps\cite{Palomaki_Nat_13}, ponderomotive squeezing\cite{Purdy_PRX_13}, and optomechanical entanglement\cite{Ockloen_Nat_18}. 
These influential demonstrations all exhibit a coherent interaction that dominates over thermal decoherence, hence satisfying the criterion $C_0 n_p>n_m$, where $n_p$ is the number of intracavity photons and $n_m$ is the number of thermal phonons.
However, Brillouin systems typically do not satisfy this criterion, primarily due to the inability to access high cooperativities together with low thermal occupation. As a result, these quantum techniques have had significantly less exposure in the Brillouin community. The system proposed here could circumvent many of these issues and, combined with the non-reciprocal nature of the interaction, may enable novel ways to produce, process and distribute quantum resources for next-generation technologies\cite{Stannigel_PRA_11, Safavi_NJP_11}.

When comparing to other Brillouin systems, it can be instructive to consider the performance of superfluid helium as a material for Brillouin scattering, independently of its desirable mechanical properties such as absence of viscosity. Superfluid helium has an extremely low optical permittivity, with $\varepsilon_{\mathrm{sf}} = n_{\mathrm{He}}^2 = 1.058$. This weak value reduces the electrostrictive forces exerted by the light~\cite{baker_photoelastic_2014}. This penalty is evidenced by the reduced electrostrictive coefficient $\gamma_e$ (Eq. (\ref{Eqelectrostrictivecoefficient})), and thus by the small value taken by the $(\varepsilon_{\mathrm{sf}} - 1)$ prefactor in Eq.~(\ref{Eq_g_0es1}). On the other hand, this is offset by superfluid helium's large compliance, which enables large strain and pressure excursions ($\tilde{\epsilon}_{v} \left(\vec{r}\right)$ term in Eq.~(\ref{Eq_g_0es1})) in response to a given force. To compare the relative magnitude of these two effects, we can consider the permittivity fluctuation $\delta \varepsilon$ per unit strain energy $u$.  This is given by $\delta \varepsilon=\sqrt{\frac{2\,u}{K}} \gamma_e$, leading to a $\gamma_e/\sqrt{K}$ figure of merit, which can be compared for different materials. This figure of merit is for instance close to three times larger for superfluid helium than for a perfectly isotropic material with silica's refractive index and bulk modulus. While this comparison is overly simplified (the tensor nature of the electrostrictive coefficient must be taken into account for solids), it indicates that superfluid helium's low permittivity is well offset by its large mechanical compliance.

\section{Conclusion}

In conclusion, we have proposed a travelling wave Brillouin platform that is based on a superfluid helium-filled slot waveguide ring resonator design. By strongly confining and co-localising optical and acoustic fields into a micrometer-cubed mode volume, this approach enables large optomechanical coupling rates and single photon cooperativities potentially exceeding unity. Such a device may be used for applications ranging from non-reciprocal optical circuits\cite{Kittlaus_NatPhot18} and superfluid based rotation sensing \cite{Lai_NatPhot_20} to interfacing with electrons on superfluid\cite{koolstra_coupling_2019}. Furthermore, this platform will enable fundamental investigations into superfluid physics with coupling to quantized vortices \cite{sachkou_coherent_2019, forstner_modelling_2019} and rotons \cite{rybalko_resonance_2007}.

\section*{Funding}
Army Research Office (W911NF17-1-0310); Australian Research Council (CE170100009,  FT140100650, DE19010031).

\section*{Acknowledgements}
This work was funded by the US Army Research Office through grant number W911NF17-1-0310 and the Australian Research Council Centre of Excellence for Engineered Quantum Systems (EQUS, project number CE170100009). W.P.B. and C.G.B respectively acknowledge Australian Research Council Fellowships FT140100650 and DE190100318.

\section*{Disclosures}
The authors declare no conflicts of interest.

\bibliography{references}

\end{document}